\documentclass[a4paper,11pt,oneside]{article}

\usepackage[top=1in,bottom=1in]{geometry}
\usepackage[T1]{fontenc}
\usepackage{parskip}
\usepackage{latexsym,amsmath,amssymb}
\usepackage{times}
\usepackage{graphicx}
\usepackage{float}
\usepackage{authblk}
\usepackage{graphicx} 
\usepackage{hyperref} 
\usepackage{amsmath}
\usepackage{amssymb}
\usepackage{amsfonts}
\usepackage{blkarray}
\usepackage{booktabs}    
\usepackage{siunitx}     
\usepackage{amssymb}
\usepackage{csquotes}
\usepackage{array}
\usepackage{gensymb}
\usepackage{bbm}
\usepackage{algorithm,algpseudocode}
\usepackage{mathtools}
\usepackage{algpseudocode}
\usepackage[table]{xcolor} 
\usepackage{colortbl} 
\usepackage{geometry} 
\usepackage[backend=bibtex,
            style=numeric,
            firstinits=true,
            maxnames=6,
            sortcites=true,
            sorting=none,
            doi=false,
            url=false,
            natbib=true,
            eprint=false,
            isbn=false]{biblatex}
\title{A Scalable Exponential Random Graph Model: Amortised Hierarchical Sequential Neural Posterior Estimation with Applications in Neuroscience}

\author[1,*]{Yefeng Fan}
\author[1,2,**]{Simon Richard White}

\affil[1]{MRC Biostatistics Unit, University of Cambridge, Cambridge, UK}
\affil[2]{Department of Psychiatry, University of Cambridge,
  Cambridge, UK}
\affil[*]{ORCID: \url{https://orcid.org/0009-0009-6039-2994}}
\affil[**]{ORCID: \url{http://orcid.org/0000-0001-8642-7037}}

\begin{document}
\maketitle

\section*{Abstract}
Exponential Random Graph Models (ERGMs) are an inferential model for analysing statistical networks. Recent development in ERGMs uses hierarchical Bayesian setup to jointly model a group of networks, which is called a multiple-network Exponential Random Graph Model (MN-ERGMs). MN-ERGM has been successfully applied on real-world resting-state fMRI data from the Cam-CAN project to infer the brain connectivity on aging. However, conventional Bayesian ERGM estimation approach is computationally intensive and lacks implementation scalability due to intractable ERGM likelihood. We address this key limitation by using neural posterior estimation (NPE), which trains a neural network-based conditional density estimator to infer the posterior.\\
We proposed an Amortised Hierarchical Sequential Neural Posterior Estimation (AHS-NPE) and various ERGM-specific adjustment schemes to target the Bayesian hierarchical structure of MN-ERGMs. Our proposed method contributes to the ERGM literature as a very scalable solution, and we used AHS-NPE to re-show the fitting results on the Cam-CAN data application and further scaled it up to a larger implementation sample size. More importantly, our AHS-NPE contributes to the general NPE literature as a new hierarchical NPE approach that preserves the amortisation and sequential refinement, which can be applied to a variety of study fields.

\section{Introduction}
Exponential random graph models (ERGMs) are statistical inference models that characterise the distribution of an observed network by flexibly using a set of network summary statistics. Standard ERGMs are one-to-one models, meaning ERGM can only characterise the topological structure of one network, and such a feature limits the application of ERGMs. More recently, a multiple-network exponential random graph model (MN-ERGM) is proposed to jointly characterise a group of networks by incorporating ERGMs into a Bayesian hierarchical structure, which shows advantages of summarising group network topology inferences as well as accounting for heterogeneity across networks \citep{Lehmann665398}. The developed MN-ERGM demonstrate strong performance in modelling neuro-imaging data, which typically requires multiple sample analysis \citep{Lehmann665398}.\\
Despite the advantage of ERGM, its estimation procedure is an extremely computationally intensive procedure in the Bayesian setup due to the sampling distribution intractability, requiring a prolonged process of exploring the mapping between the parameter space and the graph space \citep{caimo2011bayesian, caimo2017bergm}. The estimation complexity is approximated to increase exponentially with the size of the network, placing a practicality constraint on the size of the network we can fit for ERGM (usually a few hundred vertices as the maximum) \citep{complexity1, complexity222}. Such computational burden is also inherent in the MN-ERGM, and such computation demand scales linearly with the sample size. A more efficient approach is needed to scale up this advanced ERGM framework to meaningfully infer on larger samples.\\
Likelihood-free Bayesian inference offers a promising direction, which has been adopted in various scenarios where the likelihood functions are typically intractable but are possible for simulation \citep{leclercq2018bayesian, ishida2015cosmoabc}. Cranmer et al. have concluded three crucial criteria for a simulation-based approach: 1) Sampling efficiency; 2) Quality of inference; 3) Amortisation \citep{Cranmer_2020}. While the first two criteria are self-explanatory, amortisation focuses on the cost of re-fitting the model upon inference on a new observation, and it is undesirable for re-fitting to repeat computationally expensive steps. The amortisation criterion becomes especially important for the MN-ERGM application, where we need to repetitively infer on a large set of network samples.\\
More recently, a class of likelihood-free approaches called the Neural Posterior Estimation (NPE) is proposed and applied to various studies \citep{papamakarios2016fast, greenberg2019automatic, dax2021real, papamakarios2019sequential}, which trains a neural network-based conditional estimator to infer the Bayesian posterior distribution \citep{papamakarios2016fast}. NPE shows better amortisation and can flexibly model high-dimensional parameter space using expressive neural network architectures such as normalising flows, leading to more accurate and scalable likelihood-free inference \citep{pesonen2023abc, avecilla2022neural}. NPE has since been applied to a range of studies including cosmology and neuroscience \citep{fengler2021likelihood, vasist2023neural, gonccalves2020training}. Furthermore, different NPE variants have been proposed \citep{ward2022robust}, and a well-rounded package has been developed \citep{tejero2020sbi}. Among the NPE variants, a particularly important extension is the sequential neural posterior estimation (SNPE), which iteratively refines the conditional estimator towards the target posterior \citep{greenberg2019automatic}.\\
In the past, a well-rounded NPE and SNPE implementation on ERGMs builds the strong foundation for this study \citep{selfciteme}, where a systematic analysis and assessment has been performed to justify the appropriateness of NPE on ERGMs and ERGM-specific challenges we may face upon adopting NPE frameworks.\\
In this study, we propose a novel NPE extension to target a Bayesian hierarchical structure, enabling scalable implementation of NPE on MN-ERGMs. We utilise NPE to estimate the local parameters for individual networks and use an analytically tractable variational approximation update scheme for the higher hierarchical parameters \citep{jia2023analytically}. We adopt an expectation-maximisation scheme to iteratively train, refine and adjust our normalising flow-based estimator and our training data. Similar to SNPE, this allows better specification of parameters surrounding the target posteriors. Furthermore, our proposed approach preserves the amortisation properties for scalability, meaning we can incorporate a large number of targeted observed data (networks) without significantly increasing the computational complexity. In particular, we propose specific model fitting procedures that are designed to target the difficulties in using NPE-based models on ERGM density summarised from past study \citep{selfciteme}. We call our proposed approach Amortised Hierarchical Sequential Neural Posterior Estimation (AHS-NPE).\\
Apart from the specific usage of MN-ERGMs, our proposed AHS-NPE also contributes to the general NPE literature and can be adopted in other study fields. In past studies, NPE-based approaches targeting Bayesian hierarchical structures are under-explored. We now review some of those approaches. Very recently, Habermann et al. and Heinrich et al. proposed two frameworks with sufficient scalability and flexibility \citep{habermann2024amortized, heinrich2023hierarchical}. However, in comparison, our AHS-NPE has the advantage of incorporating iterative refinements of training, enabling more focus on the target posterior region. Our proposed model fulfils the future work pointed out in Habermann's model by incorporating sequential neural methods into the hierarchical modelling. A framework that can flexibly model with increasingly refined rounds is the hierarchical neural posterior estimation (HNPE) proposed by Rodrigues et al. \citep{rodrigues2021hnpe}. HNPE uses two normalising flows and aggregation networks (e.g., DeepSets) to jointly learn posterior approximations for local and global parameters. In comparison, AHS-NPE may be perceived as a specialised variant of HNPE, targeting a different class of Bayesian hierarchical frameworks. In addition, compared to HNPE, AHS-NPE preserves the amortisation property for local parameters, ensuring scalability to large datasets of observed values without introducing additional computational overhead to the training of density estimators. Tran et al. proposed a Likelihood-Free Variational Inference (HIM-LFVI) \citep{tran2017hierarchical}. HIM-LFVI shares the same objective function as our model, and therefore, our model forms an extension of HIM-LFVI. We improve by using SNPE for the local parameters without the need to perform ratio estimation. Furthermore, our AHS-NPE is more flexible in making adjustments across training rounds, allowing necessary changes to better target the ERGM densities.\\
To demonstrate our proposed AHS-NPE in this study, we implement the AHS-NPE on MN-ERGM with the real-world fMRI datasets from the Cam-CAN project \citep{shafto2014cambridge}. We perform two implementation schemes. Firstly, we aim to re-show and compare with the fitting results on the young group in Lehmann's implementation of MN-ERGM on the Cam-CAN data with 100 samples \citep{Lehmann665398}, which demonstrates our methods' reliability and estimation accuracy. Secondly, we utilise the amortisation power of our AHS-NPE and scale up the implementation of MN-ERGM on the Cam-CAN data by modelling all 586 network samples in the Cam-CAN project into two age groups, which assesses and explores the aging impact on brain connectivity. We compare and consolidate the neurological findings with Lehmann et al.'s implementation.

\section{Background Methodology, Data Definition and Problem Setup}\label{c_npe_s_npe}

\subsection{Exponential Random Graph Models}

\subsubsection{Exponential Random Graph Models}
ERGM characterises its probability distribution of network $y$ through a set of summary statistics $h(y)\in \mathbb{R}^{d}$ with parameters $\theta \in \mathbb{R}^{d}$ \citep{complexity222,book2},
\begin{equation*}
  Y \sim \text{ERGM}(\theta) \quad \text{s.t.} \quad p(Y=y|\theta)=\dfrac{\exp(\theta h(y))}{c(\theta)},
\end{equation*}
where $c(\theta)$ is an intractable normalising constant,
\begin{equation*}
  c(\theta)=\sum_{y\in \mathcal{Y}} \exp(\theta h(y)).
\end{equation*}
The intractability of the normalising constant is due to the combinatorial large graph space $\mathcal{Y}$, and this makes the sampling distribution analytically intractable.\\
In the Bayesian setting, the Exchange Algorithm is used \citep{caimo2011bayesian, caimo2017bergm}, which iteratively explores the parameter space and is computationally intensive \citep{complexity1, complexity222}. The fitting algorithm is in the Supplementary Materials.\\
Instead of modelling the network $y$ directly, past study has successfully demonstrated the appropriate usage of NPE on ERGM by modelling the summary statistics $x=h(y)$ using the fact that these statistics are Bayesian sufficient for ERGM densities \citep{selfciteme}. We define $x_{obs}$ to be the observed network statistics.\\
Importantly, to enable comparison with Lehmann's model \citep{Lehmann665398}, the summary statistics we use in this paper are the number of edges, GWESP, and GWNSP, where the exact definition can be found in the Supplimentary Materials. Further to match with the Lehmann's study, we use the same sized networks with 90 vertices \citep{Lehmann665398}, this network size is also successfully been explored in past NPE study on ERGM, giving the most consistent application performance \citep{selfciteme}.

\subsubsection{Multiple-Network Exponential Random Graph Models and Problem Setup}
We first explain the problem setup of this paper and then incorporate the concept of MN-ERGMs.\\
In our targeted Bayesian hierarchical framework, we consider a population of $n$ independent random variables, each denoted as $x_i$ for $i=1,…,n$. The sampling distribution of these random variables is $p(x_i | \theta_i)$ with parameter $\theta_i \in \mathbb{R}^{d}$. The sampling distribution is typically intractable but allows simulations.\\
To characterise the joint probability distribution of the population of samples, we build the hierarchical prior structure such that each individual-level (or local) parameter $\theta_i$ independently and identically follows a multivariate Normal distribution:
$$\theta_i \sim \mathcal{N}(\theta_g, \Sigma_g),$$
where $\theta_g$ and $\Sigma_g$ represent the group-level (or global) mean and covariance. $\theta_g \in \mathbb{R}^{d}$ and $\Sigma_g \in \mathbb{R}^{d \times d} $.\\
The hyper-prior is specified as a Normal-inverse-Wishart distribution,
$$\pi(\theta_{g}|\Sigma_{g}) \sim \mathcal{N}(\mu_0, \dfrac{\Sigma_{g}}{\kappa_{0}}),$$
$$\pi(\Sigma_{g}) \sim W^{-1}(\Psi_0, \nu_0).$$
We denote $\Theta_{all}=\{\theta_{1},...,\theta_{n},\theta_{g},\Sigma_{g}\}$.\\
Lehmann et al. proposed MN-ERGMs to jointly model a population of networks using the above Bayesian hierarchical setup \citep{Lehmann665398}, which overcomes the disadvantage of single ERGM fitting neglecting the heterogeneities across individuals. By replacing $x_i$ with the individual network samples $y_i$ and taking the sampling distribution $p(x_i | \theta_i)$ as standard ERGM, the above hierarchical structure corresponds to the MN-ERGM under Lehmann et al.'s definition. Therefore, our problem set-up targets exactly the MN-ERGM.

\subsection{Neural Posterior Estimations and Sequential Neural Posterior Estimations}
NPE and SNPE have been shown to be accurate and appropriate with ERGM application \citep{selfciteme}.\\
NPE trains a conditional density estimator to directly infer the posterior \citep{papamakarios2016fast}. More specifically, it starts by simulating a set of $B$ data-parameter pairs $\{\theta_{b}, x_{b}\}_{b=1}^{B}$ from the proposal distribution $\tilde{p}(\theta)$ to form our training dataset $D$.
$$\theta_{b} \sim \tilde{p}(\theta),$$
$$x_{b} \sim p(x|\theta_{b}).$$
A conditional density estimator $q_{\phi}(\theta|x)$ parameterised by $\phi$ is trained using the training dataset. We mainly adopt the masked autoregressive flow (MAF) as our conditional density estimator \citep{papamakarios2017masked}.\\
The conditional density estimator is trained with the objective function,
$$\mathcal{L}(\phi) = -\mathbb{E}_{\theta, x} [ \log q_\phi(\theta | x) ],$$
which is approximated by the negative log-likelihood,
$$\mathcal{L}(\phi) \approx -\sum_{b=1}^{B}log q_{\phi}(\theta_{b}|x_{b}).$$
As $B \rightarrow \infty$ \citep{papamakarios2016fast}, the conditional density estimator converges towards a density called the proposal posterior
\begin{equation}
\label{eq:c_npe_s_snpe_base_equation}
\tilde{p}(\theta|x) = \pi(\theta|x) \dfrac{\tilde{p}(\theta)p(x)}{\pi(\theta)\tilde{p}(x)},
\end{equation}
with 
$$\tilde{p}(x)=\int_{\theta}\tilde{p}(\theta)p(x|\theta).$$
When the proposal distribution is the prior distribution (i.e., $\tilde{p}(\theta)=\pi(\theta)$), we yield the standard NPE, because trained conditional density estimator targets directly the posterior (i.e., proposal posterior equals the target posterior such that $\tilde{p}(\theta|x) = \pi(\theta|x)$). We can infer a particular observation $x_{obs}$ by $q_{\phi}(\theta|x_{obs})$. Such inference process is amortised, meaning inference can be performed on different observations $x_{obs}$ without the need to re-fit the model.\\
When the proposal distribution is different from the prior distribution, we move to the SNPE concept because the conditional density estimator targets the proposal posterior instead of the posterior of interest. SNPE uses $q_{\phi}(\theta|x)$ to target the true posterior $\pi(\theta|x)$ and considers $\tilde{q}_{\phi}(\theta|x)$ to target the proposal posterior $\tilde{p}(\theta|x)$ by defining \citep{greenberg2019automatic}:
\begin{equation}
\label{eq:c_npe_s_snpe_main_equation}
\tilde{q}_{\phi}(\theta|x)=q_{\phi}(\theta|x)\dfrac{\tilde{p}(\theta)}{\pi(\theta)}\dfrac{1}{Z_{\phi}(x)}.
\end{equation}
The $Z_{\phi}(x)$ is the normalising constant,
$$Z_{\phi}(x)=\int_{\theta} q_{\phi}(\theta|x)\dfrac{\tilde{p}(\theta)}{\pi(\theta)}.$$
With the objective function,
$$L(\phi)=-\sum_{b=1}^{B} log \tilde{q}_{\phi}(\theta_{b}|x_{b}).$$
SNPE enables direct \enquote{read off} the correct conditional density estimator targeting the true posterior $\pi(\theta|x)$ without the need for post-hoc solving. To guarantee the evaluability of $Z_{\phi}(x)$, an atomic loss may be used instead. Please find more details in  \citep{greenberg2019automatic,selfciteme}.\\
Such a ratio trick used in SNPE allows the usage of a proposal distribution different from the prior so long as that proposal distribution is analytically tractable. This forms a fundamentally important concept in our later proposal of AHS-NPE. In SNPE, it iteratively uses the trained conditional density estimator $q_{\phi}(\theta|x_{\text{obs}})$ as the proposal distrbution in the next round, which facilitates step by step approaching and generating more data-parameter pairs around the target posterior $\pi(\theta|x)$. Note that the data-parameter pairs from previous rounds are re-used (See \citep{greenberg2019automatic}). This ensures the training dataset is more \enquote{relevant} to the inference aim. However, such iterative procedure uses the previous round estimated posterior and causes the amortisation property being lost (i.e., SNPE needs to be retrained when inferring a new $x_{obs}$).

\section{Methodology}\label{c_NPE_s_AHVEM}

Our proposed AHS-NPE combines analytical variational inference solutions with NPE for the individual-level parameters. We follow an iterative Expectation-Maximisation (EM) procedure for updating both individual- and global-level parameters.\\
AHS-NPE approximates the joint posterior distributions using a variational distribution $q(.)$. We use $q(\Theta_{all})$ to target $\pi(\Theta_{all}|x)$.\\
In particular, we rely on the mean-field assumption such that the joint variational distribution of parameters can be factorised into a product of independent distributions \citep{tanaka1998theory,gast2017refined},
$$q(\Theta_{all})=q(\theta_{g},\Sigma_{g})\prod_{i=1}^{n}q(\theta_{i}),$$
where the $q(\theta_{g},\Sigma_{g})$ and $q(\theta_{i})$ target the posterior $\pi(\theta_{g},\Sigma_{g}|x)$ and the conditional posterior $\pi(\theta_{i}|\theta_{g},\Sigma_{g},x)$ respectively. We then can iteratively perform tractable variational updates for $q(\theta_{g},\Sigma_{g})$ and $q(\theta_{i})$. We no longer need to jointly consider all parameters, which largely help us to preserve the model amortisation and for implementation scalability.\\
We aim to minimise the Kullbcak-Leibler (KL) divergence between our estimation and the true posterior,
$$q(\Theta_{all})=argmin KL(q(\Theta_{all})||\pi(\Theta_{all}|x))=-\int_{\Theta_{all}}q(\Theta_{all})\log \dfrac{\pi(\Theta_{all}|x)}{q(\Theta_{all})}d\Theta_{all},$$
and to approach this optimisation with the unknown true posterior $\pi(\Theta_{all}|x)$, we maximise the evidence lower bound (ELBO),
$$ELBO(q)=E_{\Theta}[\log\dfrac{\pi(\Theta_{all},x)}{q(\Theta_{all})}].$$
Under the mean-field assumption, Jia et al. have shown the optimal variational distributions are \citep{jia2023analytically},
$$q^*(\theta_{g},\Sigma_{g})\propto \exp\{E_{\theta_{i}}[\log p(\theta_{i},\theta_{g},\Sigma_{g},x)]\},$$
$$q^*(\theta_{i})\propto \exp\{E_{\theta_{g},\Sigma_{g}}[\log p(\theta_{i},\theta_{g},\Sigma_{g},x)]\}.$$
Our AHS-NPE approximates the $q^*(\theta_{i})$ using NPE to better target the intractable sampling distribution and preserve the amortisation property, where the analytical optimal solution requires analytical tractability that is not applicable to ERGM densities.\\
We use the EM algorithm to iteratively train the variational distributions and thus targets the joint posterior. The E-step uses NPE to approximate $q(\theta_{i})$ and the M-step uses the analytically optimal variational updates for the global parameters. AHS-NPE is presented in Algorithm~\ref{alg:vem}. We rigorously prove the method validity in the Supplimentary Materials.

\subsection{E-Step: Neural Posterior Estimation}
The E-step estimates the posterior distribution of $\theta_{i}$ given the current estimates of the global parameters. More specifically, we update the estimated conditional posterior $\pi(\theta_{i}|\theta_{g},\Sigma_{g},x_{i})$ by $q(\theta_{i})$ for every individual. We perform the neural posterior estimation to target the intractable sampling distribution.\\
We use a conditional density estimator $q(\theta_{i})=q_{\phi}(\theta_{i}|x_{i})$, and the posterior estimation is conditioned on the prior distribution $\pi(\theta_{i}|\theta_{g},\Sigma_{g})$ where $\theta_{g}$ and $\Sigma_{g}$ are assumed to be fixed at the updates from the M-step. In particular, we use a proposal distribution $\tilde{p}(\theta_{i})$ different from the prior $\pi(\theta_{i}|\theta_{g},\Sigma_{g})$, which is important for the ERGM-specific adjustments (discuss in Section~\ref{c_npe_s_proposal_and_ergm}). Therefore, the E-step uses the ratio trick in the SNPE with atomic/non-atomic loss, and it can be considered as an SNPE with one round of training to accommodate the different proposal distribution from the (hierarchical) prior.\\

\subsection{M-Step: Analytical Variational Updates}
M-step updates the global parameters $\theta_{g}$ and $\Sigma_{g}$ based on the updated estimation of posterior distributions $q(\theta_{i})$ for $\theta_{i}$. Jia et al. have derived an analytical solution without incorporation of any hyper-prior terms (assumed as uniform) \citep{jia2023analytically}. To fit within the context of our MN-ERGM hyper-prior setup, we derive the analytically optimal variational updates under a Normal-inverse-Wishart hyper-prior distribution for $\theta_{g}$ and $\Sigma_{g}$.
From the E-step updated $q(\theta_{i})=q_{\phi}(\theta_{i}|x_{i})$, we obtain expectation values that we use for the M-step updates,
$$\mu_i=\mathbb{E}_{\theta_i}[\theta_i],$$
$$\Sigma_i = \mathbb{E}_{\theta_i}\left[ (\theta_i - \mu_i)(\theta_i - \mu_i)^\top \right],$$
where the expectation is with respect to $q(\theta_{i})$, and we define
$$\bar{\mu}=\dfrac{1}{n}\sum_{i=1}^{n} \mu_i.$$
Analytical evaluation is not possible using NPE, and therefore, we estimate via a sampling procedure by generating $S$ samples $ \{\theta_i^{(s)}\}_{s=1}^S $ from $q_{\phi}(\theta_{i}|x_{i})$. We compute:
$$\hat{\mu_i} = \frac{1}{S} \sum_{s=1}^S \theta_i^{(s)},$$
and
$$\hat{\Sigma_i} = \frac{1}{S} \sum_{s=1}^S (\theta_i^{(s)} - \hat{\mu_i})(\theta_i^{(s)} - \hat{\mu_i})^\top.$$
We then update the analytically optimal variational distribution by
$$q^*(\Sigma_g) = W^{-1}(\Sigma_g \mid \Psi_n, \nu_n),$$
and
$$q^*(\theta_g \mid \Sigma_g) = \mathcal{N}(\theta_g \mid \mu_n, \dfrac{\Sigma_g}{\kappa_n}),$$
where
$$\nu_{n} = \nu_0 + n,$$
$$\Psi_{n} = \Psi_0 + \sum_{i=1}^{n} \left[ \Sigma_i + (\mu_i - \bar{\mu})(\mu_i - \bar{\mu})^{\top} \right] + \frac{\kappa_0 n}{\kappa_0 + n} (\mu_0 - \bar{\mu})(\mu_0 - \bar{\mu})^\top,$$
$$\mu_{n} = \frac{\kappa_0 \mu_0 + \sum_{i=1}^{n}\mu_{i}}{\kappa_0 + n},$$
$$k_{n} = \kappa_0 + n,$$
where $\nu_{n}$, $\Psi_{n}$, $\mu_{n}$ and $k_{n}$ represent the updated hyperparameters. Given the updated variational distribution representing the estimated posterior distribution of the global parameters, we need to obtain a point estimate feeding as a fixed prior structure for the E-step in the next iteration. With the closed-form variational distribution, we use the maximum a posteriori probability (MAP) such that we set
$$\Sigma_g=\dfrac{\Psi_n}{\nu_n+d+1},$$
$$\theta_g=\mu_n.$$

\begin{algorithm}[H]
\caption{AHS-NPE Fitting Process}
\label{alg:vem}

\textbf{Initialise:}
\begin{itemize}
    \item Initialise global parameters $\theta_g^{(1)} = \mu_0$, $\Sigma_g^{(1)} = \Psi_0$.
    \item Initialise proposal distribution $\tilde{p}(\theta_i)$.
    \item Initialise iteration indicator $t=1$.
\end{itemize}

\textbf{E-step (NPE):} For $i = 1, \dots, n$:\\

\begin{itemize}
    \item Simulate data-parameter pairs and argument training dataset $D$ using Algorithm~\ref{alg:c_npe_ahs_npe_proposal_update} or \ref{alg:c_npe_ahs_npe_proposal_ergm_specific}
    
    \item Train the (one-round) SNPE $q_{\phi}^{(t)}(\theta_i)$ using $D$ with prior $\pi(\theta_i \mid \theta_g^{(t)}, \Sigma_g^{(t)})$ and proposal distribution $\tilde{p}(\theta_i)$.
\end{itemize}

\textbf{M-step (Analytical Variational Updates):}\\

\begin{itemize}
    \item From $q_{\phi}^{(t)}(\theta_i)$, compute:
    $$\mu_i^{(t)} = \mathbb{E}_{\theta_i}[\theta_i],$$
    \[
    \Sigma_i^{(t)} = \mathbb{E}_{\theta_i} \bigl[ (\theta_i - \mu_i^{(t)})(\theta_i - \mu_i^{(t)})^\top \bigr],
    \]
    \[
    \bar{\mu}^{(t)} = \frac{1}{n} \sum_{i=1}^n \mu_i^{(t)}.
    \]
    \item Update variational distribution on $\theta_g$ and $\Sigma_g$:
    \[
    q^{*(t)}(\Sigma_g) = W^{-1}(\Sigma_g \mid \Psi_n, \nu_n),
    \]
    \[
    q^{*(t)}(\theta_g \mid \Sigma_g) = \mathcal{N}\bigl(\theta_g \mid \mu_n, \frac{\Sigma_g}{\kappa_n}\bigr),
    \]
    where:
    \[
    \nu_n = \nu_0 + n,
    \]
    \[
    \Psi_n = \Psi_0 + \sum_{i=1}^n \bigl[ \Sigma_i^{(t)} + (\mu_i^{(t)} - \bar{\mu}^{(t)})(\mu_i^{(t)} - \bar{\mu}^{(t)})^\top \bigr] + \frac{\kappa_0 n}{\kappa_0 + n} (\mu_0 - \bar{\mu}^{(t)})(\mu_0 - \bar{\mu}^{(t)})^\top,
    \]
    \[
    \mu_n = \frac{\kappa_0 \mu_0 + \sum_{i=1}^n \mu_i^{(t)}}{\kappa_0 + n},
    \]
    \[
    \kappa_n = \kappa_0 + n.
    \]
    \item Update prior distribution:
    \[
    \Sigma_g^{(t+1)} = \frac{\Psi_n}{\nu_n + d + 1},
    \]
    \[
    \theta_g^{(t+1)} = \mu_n.
    \]
    \item Update proposal distribution $\tilde{p}(\theta_i)$ using Algorithm~\ref{alg:c_npe_ahs_npe_proposal_update} or \ref{alg:c_npe_ahs_npe_proposal_ergm_specific}.
    
    \item $t=t+1$
\end{itemize}

\textbf{Repeat until convergence.}
\end{algorithm}

\subsection{Proposal Definition and ERGM Specific Adjustments}\label{c_npe_s_proposal_and_ergm}
\subsubsection{Proposal Definition}
The ability to extend the NPE-based framework to a large number of data points (network samples) relies on the amortisation property. Unfortunately, conventional SNPE loses the amortisation property due to the iterative revision of the proposal distribution using the fitted conditional density estimators across rounds. Therefore, the design of the proposal distribution becomes important to both preserve the amortisation across EM rounds and maintain flexibility to target ERGM specific issues.\\
The usage of atomic loss and SNPE density trick enables us to validly and flexibly choose any proposal distribution which does not need to equal the prior distribution. To construct a proposal distribution, we have the following aims and requirements.
\begin{itemize}

\item \textbf{Tractability and Evaluability}: The proposal distribution needs to be evaluatable, because we need to determine its density during the training process using atomic/non-atomic loss.

\item \textbf{Reusability and Adaptability}: The proposal should allow us with sufficient flexibility to refine our proposals in each iteration. The proposal must also be round-by-round focusing more closely on the target posterior domain. Furthermore, the proposal should re-use data from the previous training rounds.

\item \textbf{Coverage}: The proposal coverage should be sufficiently wide to avoid leakage \citep{greenberg2019automatic,tejero2020sbi,deistler2022truncated}. Importantly, it should also consider correct coverage of the prior and possible posterior domains.

\item \textbf{Adjustment Flexibility}: The proposal should allow flexible adjustments across iterations. It is useful for our next ERGM-specific adjustments.

\end{itemize}

We now design a proposal and its proposal update scheme to jointly satisfy those aims. To constrain the proposal distributions on parametrised densities and take advantage of the Normal nature of the prior distribution, we use a mixture of Normal as the proposal distribution such that:
$$\tilde{p}(\theta) = \sum_{t} w_t \, \mathcal{N}(\theta | \theta_g^{(t)}, \Sigma_g^{(t)}),$$
where $t$ is the AHS-NPE round iterations. Importantly, we iteratively incorporate (and may also discard previous) updated hierarchical prior $\mathcal{N}(\theta | \theta_g^{(t)}, \Sigma_g^{(t)})$ for individual-level variables into the proposal. This has several significant benefits. Firstly, it ensures (at least minimal) coverage on the prior, which is important to avoid leakage issues \citep{deistler2022truncated}.\\
Moreover, it enables reuse of samples from previous rounds. Practically, we generate $N_t$ data-parameter pairs based on the revised prior distribution in each iteration. These are consequently combined with training data from previous rounds to yield a total of $N$ data-parameter pairs. We estimate the mixture weights by $w_t = \frac{N_t}{N}$ such that the proposal distribution unbiasedly represents the joint training dataset\footnote{The mixture weights need to be re-adjusted in each AHS-NPE round, to ensure the weights are representative and sum to one.}. Across iterations, we may also discard data from \enquote{distant} iterations and maintain the overall sample size using only the most \enquote{relevant} data-parameter pairs.\\
Furthermore, our proposal distribution does not implicitly depend on the trained conditional density estimator inferred on any observed data, while still maintaining the ability to iteratively refine the model. This differentiates it from the SNPE, and our proposal preserves the amortisation to enable scalability on a large set of observed data (networks).\\
Finally, the variational EM scheme iteratively refines the prior distributions on local parameters. This means every trained density estimator across iterations is independent. This gives us the flexibility to augment the neural-network architecture, types of density estimators, data-parameter training datasets, and proposal distributions across iterations. Such flexibility proves to be especially important for us to make ERGM-specific adjustments.\\
The designed proposal distribution and proposal update scheme fulfills all the aims we want to target; we include a general proposal update scheme in Algorithm~\ref{alg:c_npe_ahs_npe_proposal_update}. In addition, if the density estimator is chosen as MDN, we can more easily adopt an SNPE-A scheme \citep{papamakarios2016fast}, which does not need to use atomic loss and can better avoid leakage issues (see \citep{greenberg2019automatic} and Supplementary Material in \citep{selfciteme}).

\begin{algorithm}[H]
  \caption{AHS-NPE Proposal Update Scheme}
  \label{alg:c_npe_ahs_npe_proposal_update}
  \textbf{Initialise:}
  \begin{itemize}
  \item Set initial proposal as $\tilde{p}_1(\theta)$.
  \item Set initial sampling component the same as local parameter prior, $\tilde{p}_1(\theta)=\tilde{p}^{(1)}(\theta) = \mathcal{N}(\theta | \theta_g^{(1)}, \Sigma_g^{(1)})$.
  \item Set initial training dataset $D = \emptyset$.
  \item Set $N = 0$.
  \end{itemize}
  
  \textbf{For each $t = 1$ to $T$:}
  \begin{itemize}
  \item Sample $\theta_{b,t} \sim \tilde{p}^{(t)}(\theta)$ for $b = 1, \dots, N_{t}$.
  \item Simulate $x_{b,t} \sim p(x|\theta_{b,t})$ for $b = 1, \dots, N_{t}$.
  \item Aggregate $D = D \cup \{x_{b,t},\theta_{b,t}\}^{N_{t}}_{b=1}$ and update $N = N + N_{t}$.
  \item Update all weight component $w_{t} = \frac{N_t}{N}$ and $\sum_{t} w_{t}=1$.
  \item Perform AHS-NPE update using Algorithm~\ref{alg:vem} with training dataset $D$ to obtain updated prior (on local parameters) $$\mathcal{N}(\theta | \theta_g^{(t)}, \Sigma_g^{(t)}).$$
  \item Update proposal component $$\tilde{p}^{(t+1)}(\theta) = \mathcal{N}(\theta | \theta_g^{(t)}, \Sigma_g^{(t)}).$$
  \item Update proposal distribution $$\tilde{p}_{t+1}(\theta) = \tilde{p}_{t}(\theta) + w_{t+1} \, \mathcal{N}(\theta | \theta_g^{(t)}, \Sigma_g^{(t)}).$$
  \end{itemize}
\end{algorithm}

\subsubsection{ERGM Specific Adjustments}
Previous successful implementations of both NPE and SNPE on ERGMs have summarised a series of ERGM-specific challenges for NPE and SNPE usage  \citep{selfciteme}, and we summarise them in the following as the aspects we want to target and resolve. These summaries are highly condensed; for more comprehensive explanations and additional details, please refer to \citep{selfciteme}. Taking advantage of the flexibility of AHS-NPE, we propose a distinct training process to address those ERGM-specific issues.

\begin{itemize}

\item \textbf{Boundary Effect}: As explored in the past study \citep{selfciteme}, the \enquote{boundary effect} leads to sharp transitions or plateaus between the data-parameter mapping, for which a very expressive density estimator may overfit those \enquote{noises} and cause fitting or leakage issues \footnote{Leakage refers to when the trained density estimator place a lot of probability mass to the incorrect regions of the parameter space under the true posterior distribution.}, especially when the coverage is large.

\item \textbf{Multi-modality}: Depending on the choice of summary statistics, ERGMs are likely to exhibit multi-modal issues (the same set of parameters characterising multiple graph space domains through ERGM densities), which can compromise the training data quality.

\item \textbf{Wide Coverage}: An initial wide coverage should be adopted to offer support for various probable posterior domains for different observations. ERGM statistics inter-relation may mask the diagnostic of the incorrectly estimated posterior due to narrow coverage \citep{selfciteme}. This is also especially important for our AHS-NPE application when there is large sample heterogeneity.

\item \textbf{Not Too Wide Coverage}: We should also avoid excessively wide proposals, which may contain various parameter space domain that exhibits boundary effects and multi-modalities. In particular, existence of boundary effects may restrict the adoption of more expressive density estimators as they may cause over-fitting and leakage.

\end{itemize}

To target the above issues, we propose an ERGM-specific adjustment scheme. To ensure sufficient coverage, we set our initial proposal distribution as a wide Normal distribution $\mathcal{N}(\theta | \mu_{\text{initial}},\Sigma_{\text{initial}})$ with an increased sampling size $N_{1}$ (i.e., $N_{1}>N_{t}$ for $t=1,...,T$).\\ 
Given a sufficiently wide initial proposal component and proposal distribution, the risk of \enquote{boundary effect} is triggered. To address this, we consider a burn-in period for density estimator training. During the burn-in period, we use a less expressive density estimator $q^{\text{initial}}_{\phi}(\theta|x)$ up to iteration $T_{\text{initial}}$ to avoid over-fitting to the \enquote{boundary effect}.\\
Finally, when we iteratively learn more about the location of the \enquote{true} posteriors during our burn-in period, we then focus on that posterior domain and target the multi-modal issues within that domain. To achieve that, we switch to a more expressive and deeper density estimator $q_{\phi}(\theta|x)$ for iterations after burn-in. Furthermore, we discard the wide first-round data-parameter pairs $D_{\text{initial}}$ from the training dataset. It is then replaced by a more refined $D_{\text{refined}}$ to ensure adequate coverage without including unnecessarily wide parameter space that may exhibit boundary effects. More specifically, the $D_{\text{refined}}$ is drawn from a Normal distribution $\mathcal{N}(\theta | \mu_{\text{refined}}, \Sigma_{\text{refined}})$ with size $N_{\text{refined}}$, incorporating all information learned from the burn-in rounds. This adjustment ensures that we can use more expressive density estimators without being adversely affected by the noisy wide coverage. It also ensures that the updated training dataset provides an adequate coverage to avoid leakage issues using the atomic loss. We summarise our ERGM specific adjustment proposal update scheme in Algorithm~\ref{alg:c_npe_ahs_npe_proposal_ergm_specific}.\\
More specifically, in our AHS-NPE application, we set
$$\mathcal{N}(\theta | \mu_{\text{initial}}, \Sigma_{\text{initial}})=\mathcal{N}(\theta | 0, 10I),$$
ensuring sufficiently wide initial proposal support to cover all possible posterior distributions. Furthermore, we set $N_1=100,000$, $N_t=20,000$ and $N_{\text{refined}}=50,000$. For the refined proposal component after the burn-in period, $\mathcal{N}(\theta | \mu_{\text{refined}}, \Sigma_{\text{refined}})$, we average the proposal component means and covariances across burn-in rounds and inflate the covariance by a factor of 5 times to improve coverage. More specifically, we set:
$$\mu_{\text{refined}}=\dfrac{\sum_{t=2}^{T_{\text{initial}}}\theta_g^{(t)}}{T_{\text{initial}}-1},$$
$$\Sigma_{\text{refined}}=5\dfrac{\sum_{t=2}^{T_{\text{initial}}}\Sigma_g^{(t)}}{T_{\text{initial}}-1}.$$
For the density estimators during burn-in rounds $q^{\text{initial}}_{\phi}(\theta|x)$, we use MAF with 32 hidden units (neurons) and 5 transformation steps in the flow. We only use a small number of hidden units to avoid being too expressive. We set $T_{\text{initial}}=4$. After the burn-in period (i.e., at iteration 5), we increase the density estimator $q_{\phi}(\theta|x)$ hidden units to 64 and the number of transformation steps to 10. In particular, we significantly inflate the transformation steps to increase the depth of the flow. More transforms allow for finer adjustments to the data distribution to handle extra complexity. This is especially important for modelling distributions with abrupt jumps and multi-modality.

\begin{algorithm}[H]
\caption{AHS-NPE ERGM Adjustment Scheme}
\label{alg:c_npe_ahs_npe_proposal_ergm_specific}

\textbf{Initialisation:}
    \begin{itemize}
        \item Set initial proposal as $\tilde{p}_1(\theta) = \mathcal{N}(\theta | \mu_{\text{initial}}, \Sigma_{\text{initial}})$.
        \item Set initial sampling component $\tilde{p}^{(1)}(\theta) = \mathcal{N}(\theta | \mu_{\text{initial}}, \Sigma_{\text{initial}})$.
        \item Initialise training dataset $D = \emptyset$; set $N = 0$.
        \item Set initial training data size $N_1$ such that $N_1 > N_t$ for $t = 1, \dots, T$.
    \end{itemize}

    \textbf{For each $t = 1$ to $T$:}
    \begin{itemize}
        \item Sample $\theta_{b,t} \sim \tilde{p}^{(t)}(\theta)$ for $b = 1, \dots, N_t$.
        \item Simulate $x_{b,t} \sim p(x|\theta_{b,t})$ for $b = 1, \dots, N_t$.
        \item Aggregate $D = D \cup \{x_{b,t}, \theta_{b,t}\}_{b=1}^{N_t}$ and update $N = N + N_t$.
        \item Update all weight component $w_{t} = \frac{N_t}{N}$ and $\sum_{t} w_{t}=1$.
    \end{itemize}

    \textbf{If $t = T_{\text{initial}}$:}
    \begin{itemize}
        \item Revise proposal by removing initial wide coverage data:
        \[
        \tilde{p}_t(\theta) = \tilde{p}_t(\theta) - w_1 \mathcal{N}(\theta | \mu_{\text{initial}}, \Sigma_{\text{initial}}).
        \]
        \[
        D = D \setminus \{x_{b,1}, \theta_{b,1}\}_{b=1}^{N_1}, \quad N = N - N_1.
        \]

        \item Define refined sampling component:
        \[
        \tilde{p}^{(\text{refined})}(\theta) = \mathcal{N}(\theta | \mu_{\text{refined}}, \Sigma_{\text{refined}}).
        \]

        \item Sample from the refined sampling component and update training data:
        \[
        D = D \cup \{x_{b,\text{refined}}, \theta_{b,\text{refined}}\}_{b=1}^{N_{\text{refined}}}, \quad N = N + N_{\text{refined}}.
        \]

        \item Revise proposal by incorporating refined coverage and adjust all weights:
        \[
        \tilde{p}_t(\theta) = \tilde{p}_t(\theta) + w_{\text{refined}} \mathcal{N}(\theta | \mu_{\text{refined}}, \Sigma_{\text{refined}}),
        \]
        \[
        w_{\text{refined}} = \dfrac{N_{\text{refined}}}{N}.
        \]
    \end{itemize}

    \textbf{If $t < T_{\text{initial}}$:}
    \begin{itemize}
        \item Perform AHS-NPE update using Algorithm~\ref{alg:vem} with training dataset $D$ and $q^{\text{initial}}_{\phi}(\theta|x)$.
        \item Obtain updated prior $\mathcal{N}(\theta |  \theta_g^{(t)}, \Sigma_g^{(t)}  )$.
    \end{itemize}
    
    \textbf{Else:}
    \begin{itemize}
        \item Perform AHS-NPE update using Algorithm~\ref{alg:vem} with training dataset $D$ and $q_{\phi}(\theta|x)$.
        \item Obtain updated prior $\mathcal{N}(\theta |  \theta_g^{(t)}, \Sigma_g^{(t)}  )$.
    \end{itemize}

    \textbf{Update proposal component:}
    \[
    \tilde{p}^{(t+1)}(\theta) = \mathcal{N}(\theta |   \theta_g^{(t)}, \Sigma_g^{(t)}  ).
    \]

    \textbf{Update proposal distribution:}
    \[
    \tilde{p}_{t+1}(\theta) = \tilde{p}_t(\theta) + w_{t+1} \mathcal{N}(\theta |   \theta_g^{(t)}, \Sigma_g^{(t)}  ).
    \]

\end{algorithm}

\section{Implementation Scheme}\label{c_npe_s_camcan}
Our implementation of AHS-NPE on MN-ERGM targets mainly the Cam-CAN project, which is the real-world resting-state fMRI data gathered with the research aim of investigating the aging effect on brain \citep{shafto2014cambridge}. Lehmann et al. successfully implemented the MN-ERGM on the Cam-CAN data to capture the intuitive aging impact on brain connectivity \citep{Lehmann665398}.\\
To assess and fully utilise the scalability of our AHS-NPE, we consider two implementation schemes: One for validating the usefulness and adequacy of AHS-NPE, and another one for an extended MN-ERGM implementation on Cam-CAN data.\\
For our implementations of AHS-NPE, we use the ERGM specific adjustment scheme as outlined in Algorithm~\ref{alg:c_npe_ahs_npe_proposal_ergm_specific} Section~\ref{c_npe_s_proposal_and_ergm}.\\
To detect training convergence, we monitor the relative change in global-level parameters between iterations  (i.e., $\frac{|\theta_g^{(t)} - \theta_g^{(t-1)}|}{|\theta_g^{(t-1)}|}$). We conclude convergence if the average relative changes across all parameters remain below 0.01 for two consecutive rounds.\\
As a convention, we draw 100,000 posterior samples for both the individual- and group-level parameters unless specified otherwise.

\subsection{Reproducing Cam-CAN results}

In the first implementation, we aim to reproduce the Cam-CAN data MN-ERGM implementation results from the Lehmann et al. paper \citep{Lehmann665398}. To replicate the same implementation scheme, we focus on the One-level model for the young group (Section 2.5.1. in \citep{Lehmann665398}). In particular, we target the same network set with the same data processing scheme by similarly binarising networks under absolute thresholding with group-wide average node degree set to 3, and use a total sample of 100 of the youngest individuals in the Cam-CAN data.\\
However, the hyper-prior settings differ. In Lehmann et al. implementation, they adopt the hyper prior,
$$\pi(\theta_{g}) \sim N(\mu_0,\Sigma_0),$$ 
$$\pi(\Sigma_{g}) \sim W^{-1}(\Psi_0, \nu_0),$$
which follows a Normal and Inverse-wishart distribution, but we adopt a Normal-inverse-Wishart distribution,
$$\pi(\theta_{g}|\Sigma_{g}) \sim \mathcal{N}(\mu_0, \dfrac{\Sigma_{g}}{\kappa_{0}}),$$
$$\pi(\Sigma_{g}) \sim W^{-1}(\Psi_0, \nu_0).$$
We therefore re-implement a Bayesian multi-network ERGM with our Normal-inverse-Wishart distribution. We use the same informative hyper-prior specification for $\{\mu_0,\Psi_0,\nu_0\}$ as Lehmann et al. (see details in \citep{Lehmann665398}) and set $k_{0}=1$. We refer to this fitting as \textbf{our Bayesian fitting} to distinguish it from the Lehmann et al. fitting.\\
In this first implementation scheme, we assess and compare our AHS-NPE fitting with both our Bayesian fitting and Lehmann et al. implementation results \citep{Lehmann665398}.

\subsection{Extending Cam-CAN results}
The MN-ERGM in Lehmann et al. implementation assigns the 100 youngest and 100 oldest individuals to two age groups for implementation and posterior density comparisons.\\
To extend upon that, we implement our AHS-NPE on two age groups, each with a sample size of 293, forming a total sample size of 586. Each group implementation is equivalent to the One-level model in Lehmann et al. paper \citep{Lehmann665398}. To form the age groups, we divide the oldest half of individuals into the old group and the youngest half of individuals into the young group. Networks are also binarised by absolute thresholding with group-wide average node degree set to 3. We use the same approach to specify the informative hyper-priors as Lehmann et al. for $\{\mu_0,\Psi_0,\nu_0\}$ and set $k_{0}=1$.

\section{Implementation Result}

\subsection{Reproducing Cam-CAN results}
\begin{center}
\begin{figure}[H]
\centering
\includegraphics[width=5.8in, height=5.5in]{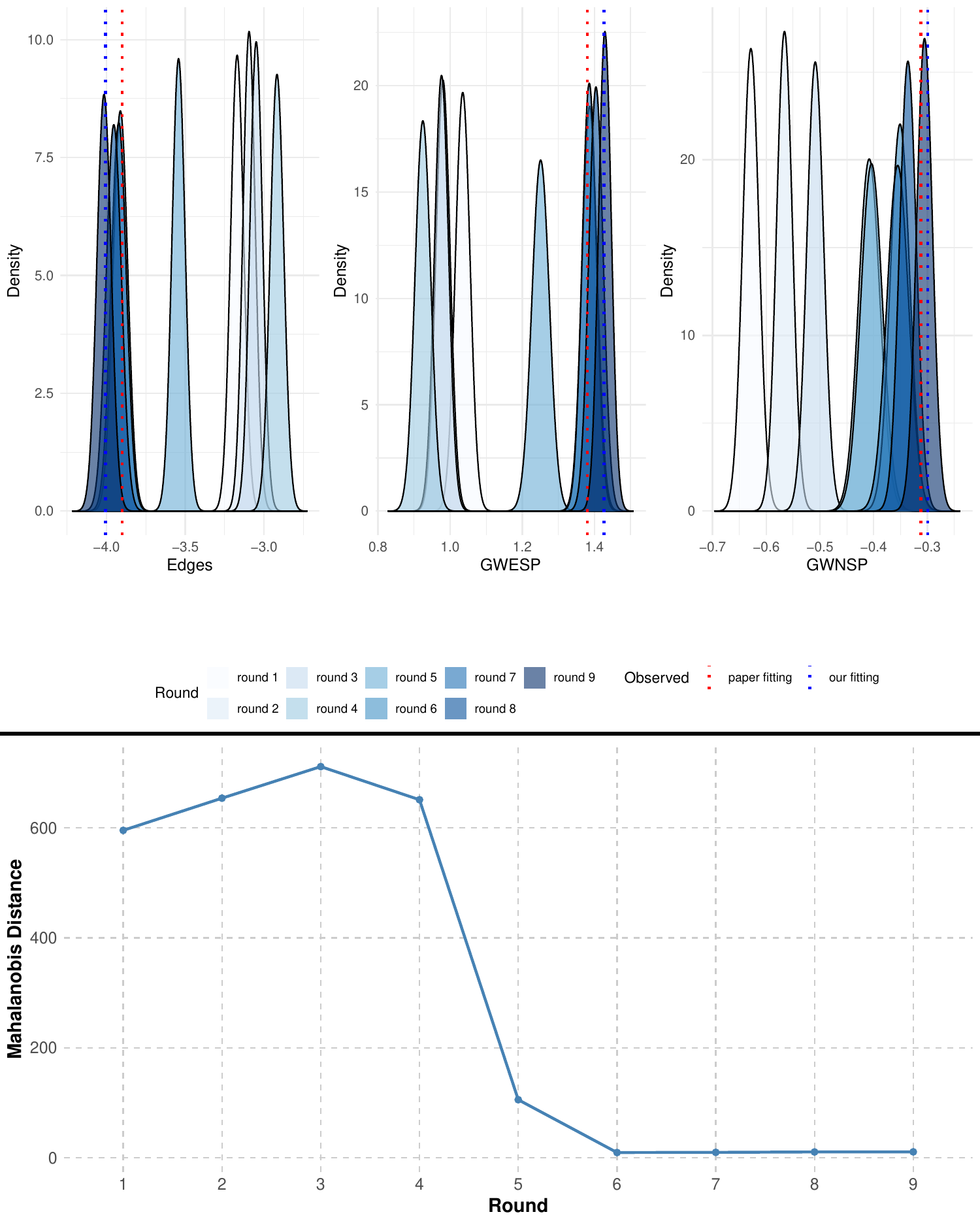}
\caption{Plots for comparing AHS-NPE to our Bayesian fitting and Lehmann et al. fitting \citep{Lehmann665398}; in the above plot, we plot posterior densities of $\theta_{g}$ in AHS-NPE training rounds; the red dashed lines are for the posterior means of $\theta_{g}$ in Lehmann et al. fitting and the blue dashed lines are for our Bayesian fitting; we only display 9 rounds due to the color spectrum limit, further rounds show almost the same density as Round 9 (see our convergence criterion); we do not have exact results from Lehmann et al. fitting, and our display is based on visual inspection of their Figure 2 in \citep{Lehmann665398}; the below plot is for the Mahalanobis distance \citep{mclachlan1999mahalanobis} between our AHS-NPE fitting on $\theta_{g}$ and the posterior means of our Bayesian fitting; for consistency with the above plot, 9 rounds are used; Mahalanobis distance is a measure of the distance between a point and a distribution \citep{mclachlan1999mahalanobis}.}
\label{fig:AHSNPE_validate_group_level}
\end{figure}
\end{center}

\begin{center}
\begin{figure}[H]
\centering
\includegraphics[width=5.5in, height=6in]{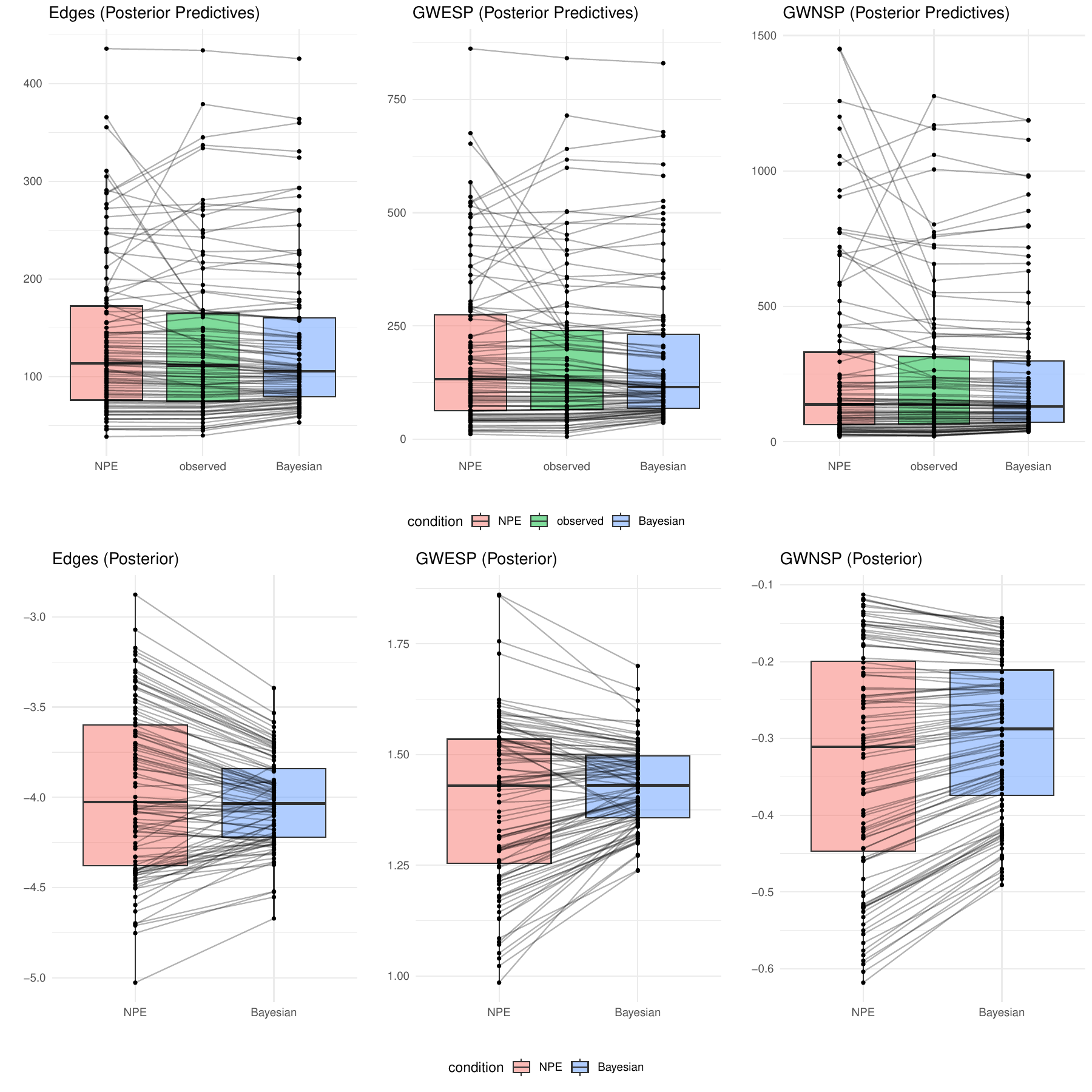}
\caption{Paired plots that compare AHS-NPE with our Bayesian fitting on the individual-level parameters $\theta_{i}$; in the above plots, we display the posterior predictive means based on individual-level realisations, and they are plotted as paired plots between AHS-NPE (red), observed statistics (green), and our Bayesian fitting (blue); in the below paired plots, we show the posterior means of $\theta_{i}$ between AHS-NPE (red) and our Bayesian fittings (blue); note that we do not have access to Lehmann et al.'s fitting results on individual-level posteriors, and therefore, we omit comparison with their fittings here.}
\label{fig:AHSNPE_validate_posterior_density_and_pp}
\end{figure}
\end{center}

\begin{center}
\begin{figure}[H]
\centering
\includegraphics[width=5in, height=5in]{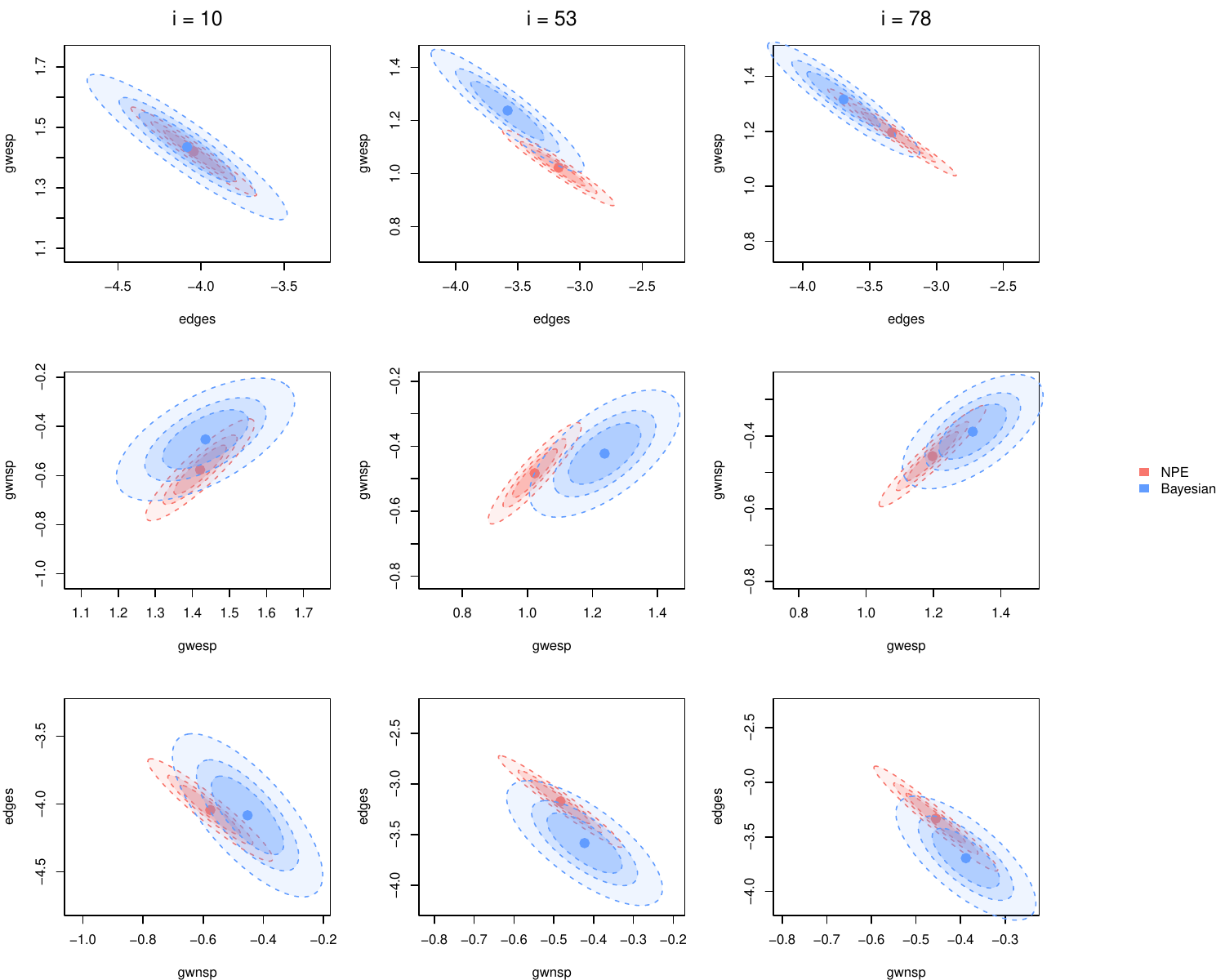}
\caption{Pairwise contour plots for posterior densities on individual-level parameters; the red is for AHS-NPE and the blue is for our Bayesian fitting; the contours are for the $50\%$, $75\%$ and $95\%$ credible intervals, respectively; we assess the 10th, 53rd and 78th sample in the young group. This matches with Lehmann et al. displayed plot in Figure 3 \citep{Lehmann665398}.}
\label{fig:AHS_NPE_validate_pairwise_individual}
\end{figure}
\end{center}

\begin{center}
\begin{figure}[H]
\centering
\includegraphics[width=5in, height=3.7in]{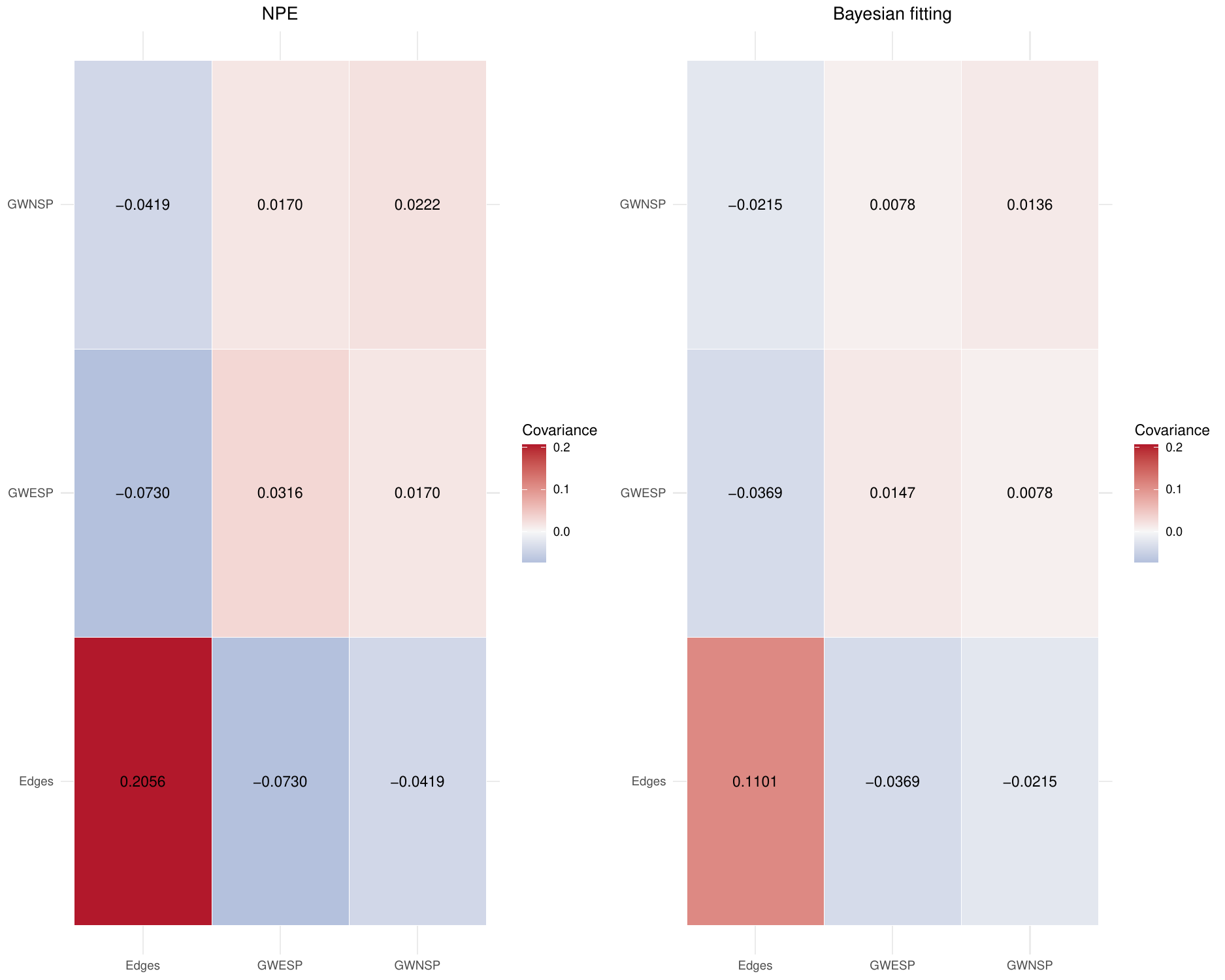}
\caption{Heat plot styled visualisation of MAP of group-level covariances $\Sigma_{g}$; the magnitudes are reflected by colours and transparencies; the exact values are displayed in black labellings accordingly.}
\label{fig:AHSNPE_validate_group_level_covariance}
\end{figure}
\end{center}

We implement the AHS-NPE on the young group, as described in Section~\ref{c_npe_s_camcan} and use the ERGM-specific adjustments (Section~\ref{c_npe_s_proposal_and_ergm}). We perform 11 iterations to achieve convergence.\\
We first assess and validate the AHS-NPE results for the group-level parameters $\theta_{g}$ by comparing them with our Bayesian fitting and Lehmann et al.’s fitting. In Figure~\ref{fig:AHSNPE_validate_group_level} (above), we see that AHS-NPE successfully converges to our Bayesian fitting by iteration 9, showing little bias on the group-level parameters, and they are similar to Lehmann et al.'s fitting on an absolute scale. We also assess the Mahalanobis distance on group-level mean between AHS-NPE and our Bayesian fitting (Figure~\ref{fig:AHSNPE_validate_group_level} below). We can see the estimations are rough in the burn-in rounds 1 to 4, but these rounds are crucial for adequate coverage for further training iterations. The estimation rapidly improves after burn-in due to refined coverage and enhanced density estimators.\\
We then assess the AHS-NPE on individual-level parameters $\theta_{i}$ by comparing them with our Bayesian fitting in Figure~\ref{fig:AHSNPE_validate_posterior_density_and_pp}. While the distribution of posterior means aligns relatively well (Figure~\ref{fig:AHSNPE_validate_posterior_density_and_pp} below), AHS-NPE shows a wider spread compared to our Bayesian fitting. However, we should not treat the Bayesian fitting as the \enquote{truth}, and therefore, we assess the posterior predictive distributions of both AHS-NPE and our Bayesian fitting with the observed statistics (Figure~\ref{fig:AHSNPE_validate_posterior_density_and_pp} above). Both AHS-NPE and Bayesian fit show adequate alignments. More specifically, the AHS-NPE shows better and more precise alignment with the observed values in most samples, whereas our Bayesian fitting shows systematic drifts in most sample predictions. However, the Bayesian fitting appears more accurate for more extreme samples than our AHS-NPE.\\
We also examine the covariances of the individual-level parameters under AHS-NPE and our Bayesian fitting (Figure~\ref{fig:AHS_NPE_validate_pairwise_individual}). While the mean values are similar between the two, AHS-NPE produces a tighter posterior credible interval. Furthermore, AHS-NPE successfully captures the intrinsic inter-relationships between summary statistics, showing similar covariance directions to those in our Bayesian fitting. Both AHS-NPE and our Bayesian fitting are relatively consistent with Lehmann's fitting (see Figure 3 in \citep{Lehmann665398}).\\
We then assess the group-level covariance $\Sigma_{g}$, which captures heterogeneities across individuals. In Figure~\ref{fig:AHSNPE_validate_group_level_covariance}, while $\Sigma_{g}$ estimated by AHS-NPE MAP and our Bayesian fitting are relatively similar and consistent in trends, AHS-NPE estimates a larger group-level covariance than our Bayesian fitting. This reflects the same situation that we observed in Figure~\ref{fig:AHSNPE_validate_posterior_density_and_pp} (below) where AHS-NPE individual-level parameters are more widely spread than our Bayesian fitting. Two reasons may be possible. Firstly, the bias in AHS-NPE individual-level estimates accumulates to the group-level covariances. Secondly, the centred design of our Bayesian fitting imposes a shrinkage effect to \enquote{tie} the individual-level parameters around the group-level parameters such that the inferred $\Sigma_{g}$ is smaller than the \enquote{true} variability (i.e., our Bayesian fitting underestimates rather than AHS-NPE overestimates). Although these two reasons may be compounding, we suspect the second reason is more likely because it is observed that our Bayesian fitting shows systematic drifts in most posterior predictive samples compared to the observed values (Figure~\ref{fig:AHSNPE_validate_posterior_density_and_pp} above).\\
Overall, AHS-NPE demonstrates accurate estimation and consistency with conventional Bayesian fittings, aligning closely with both our Bayesian fitting and Lehmann et al.'s fitting \citep{Lehmann665398}. Our AHS-NPE successfully re-show the fitting results on the Cam-CAN young group.

\subsection{Extending Cam-CAN results}\label{s_camcan_extensionssss}
\begin{center}
\begin{figure}[H]
\centering
\includegraphics[width=4in, height=5in]{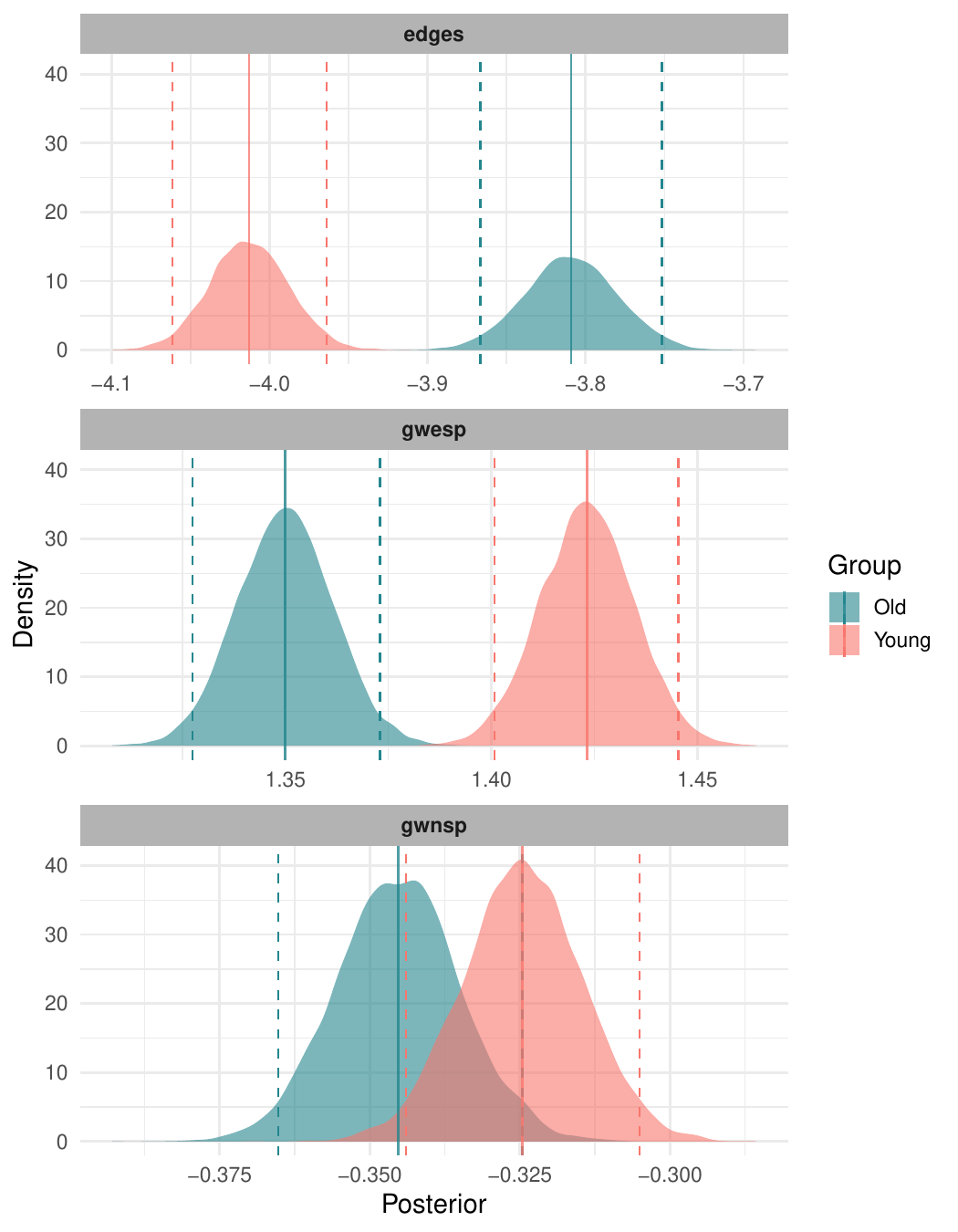}
\caption{Density plots of the posterior density estimates for group-level means; the young groups are plotted in red and the old groups are plotted in green blue (colours match with Lehmann et al. paper \citep{Lehmann665398}); three statistics are considered: edges (above), GWESP (middle) and GWNSP (below); the density plots use 10,000 posterior samples; the lines are for the $95\%$ credible intervals and posterior means.}
\label{fig:ahs_npe_young_old_group_results}
\end{figure}
\end{center}

We now implement the AHS-NPE on the Cam-CAN data to enable a larger-scale fitting with 293 samples for each age group. We reach convergence in 12 iterations for the old group and 11 iterations for the young group. We assess the group-level posterior and its $95\%$ credible interval (Figure~\ref{fig:ahs_npe_young_old_group_results}). We observe that the young group has significantly larger GWESP and GWNSP posterior than the old group, but the edge posterior is significantly smaller in the young group.\\
Large GWESP and GWNSP with a small edge indicate a very efficient system. Therefore, our results inferred for the young group means efficient brain connectivity, which corresponds to the small-world properties that have been found among young people in other studies when analysing neuroimaging networks \citep{eeg_net1,eeg_net2}. This may be because young people maintain strong white matter integrity, allowing information to travel efficiently within the brain, but such efficiency is lost during ageing.\\
Findings from our extended AHS-NPE fitting are highly consistent with the MN-ERGM fitting by Lehmann et al. \citep{Lehmann665398}. In particular, the directions (i.e., signs of posterior estimates) of their posterior estimates are consistent with our result, for which they provided a detailed discussion to yield a similar ageing and connectivity efficiency conclusion as ours (see \citep{Lehmann665398}). Moreover, our fitting shows relatively more significant age group differences. Our AHS-NPE can potentially be adopted in a much larger study to yield even more robust age difference results.\\
Overall, our AHS-NPE extends the Cam-CAN application to re-validate the fitting results from past studies, and it seems to consolidate some extra interesting age differences.

\section{Discussion}

Through our implementations, we have established the accuracy and usefulness of our novel AHS-NPE by re-showing the young cohort fitting results of MN-ERGMs in Lehmann study \citep{Lehmann665398}. Such reproducing shows our AHS-NPE is a reliable tool for network analysis studies and broader studies in other fields.\\
AHS-NPE add value to the ERGM literature by providing a much scalable modelling framework, which proves to be essential for the developed MN-ERGMs. The conventional Bayesian ERGM fitting is subject to computational burden, which is mainly due to the cost of sampling updates for the auxiliary variable (i.e., network simulation in the exchange algorithm). In our AHS-NPE implementation to extend the Cam-CAN study (Section~\ref{s_camcan_extensionssss}) to target 586 samples, with 11 iterations to reach convergence and simulate 10,000 posterior samples per individual, we performed 330,000 simulations of data-parameter pairs for the model training. To perform the same scaled implementation using conventional Bayesian fitting, we need at least 5,860,000 simulations of the auxiliary variables, which does not take into consideration the burn-in period. More importantly, the computational demand for the conditional density estimator training and data-parameter pairs simulation are almost the same between our implementation on 100 and 286 samples (per group), and this is because the amortisation property means once the conditional density estimator is trained, the inference step computation is negligible. An important implication of that is we can further scale up the implementation sample size without the need to increase the computational burden significantly.\\
Furthermore, our AHS-NPE preserves both the iterative refinement and the amortisation compared to other NPE-based hierarchical frameworks \citep{habermann2024amortized,rodrigues2021hnpe}. While our implementation focus is mainly on ERGMs, its application and usage can be extended to the broad NPE literature on hierarchical structures.

\subsection{Limitation and Future Work}\label{c_npe_s_discussion}
In the AHS-NPE, we have established well-rounded guidelines on specific problems we may face when implementing NPE on ERGMs and designed procedures accordingly to target them. We are confident that using our design procedures, we can yield an adequate AHS-NPE fitting, and it is successfully validated using our implementations. However, we cannot claim that our procedures and the identified issues are the definite \enquote{truth}, because investigating the root \enquote{true} cause in scenarios where we are likely to face ERGM-specific challenges and how our proposed strategy solves it is challenging in a neural network setup. Our established procedures rely on our trials, experiences and knowledge of ERGMs, and they are certainly useful.\\
While our AHS-NPE shows multiple advantages in terms of armortisation and iterative refinement, the generalisability of our proposed framework is relatively limited in comparison to other NPE-based hierarchical frameworks  \citep{habermann2024amortized,rodrigues2021hnpe}. The implementation logic can be generalised to any framework setup with a tractable analytical variational solution. The proposal density can be adjusted to other mixture distributions using similar logic. While the ERGM-specific adjustment scheme can also be adjusted and generalised, the refined sampling component $\tilde{p}^{(\text{refined})}(\theta)$ may require some careful design, which should aim to condense but still maintain a reasonable coverage using the burn-in rounds information. We also discussed some other potential limitations when proving methods validity in the Supplementary Materials.\\
In our Cam-CAN application in Section~\ref{s_camcan_extensionssss}, we extended the study to a larger scale. However, the group division imposes some limitations on our neurological interpretations as the age range targeted may be too broad. In other words, the distinction between an individual being young or old is too general, which may introduce bias to our findings of age-related differences. Given more available data and sufficient contextual knowledge, future implementations can better utilise the strength of AHS-NPE to target specific neuro-imaging research questions.

\section{Supplementary Material}

\subsection{Bayesian ERGM estimation}
Estimation of ERGM in the Bayesian setup uses the Exchange Algorithm \citep{caimo2011bayesian, caimo2017bergm}. It iteratively considers exchanging with an auxiliary variable to cancel out the intractable nising constants.
See Algorithm~\ref{alg:c_review_network_simulation} and \ref{alg:approximate_exchange}.

\begin{algorithm}[H]
  \caption{Network Simulation Algorithm With Given $\theta$}\label{alg:c_review_network_simulation}
  \begin{algorithmic}
    \State \textbf{Initialisation}
    \State Initialise a network $y^{(0)}$ of the same size as the targeting network\\
    \For {$i=1,...,I$}
    \State Propose $y'$ from $q(.|y^{(i-1)})$
    \State Accept $y'$ with\\
    \State $$\text{Acceptance Ratio}=\dfrac{p(y'|\theta)q(y'|y^{(i-1)})}{p(y^{(i-1)}|\theta)q(y^{(i-1)}|y')},$$\\
   
    \EndFor

  \end{algorithmic}
\end{algorithm}

\begin{algorithm}[H]
  \caption{Exchange Algorithm}
  \label{alg:approximate_exchange}
  \begin{algorithmic}
    \State \textbf{Initialise:}
    Set $T$ and initialise $\theta^{(1)}$\\
    \For{$t=1,...,T$}
    \State Propose
    $\theta_{j}' \sim \mathcal{N}(\theta_{j}^{(t)}, \Sigma)$
    \State
    Simulate $y_{j}' \sim p(\cdot|y_{j-1},\theta_{j}')$ with
    Algorithm~\ref{alg:c_review_network_simulation}
    \State Accept the
    proposal with
\begin{equation*}
\begin{aligned}
\text{Acceptance Ratio} 
&= \frac{\pi(\theta' \mid y)}{\pi(\theta \mid y)} \cdot \frac{\pi(y' \mid \theta)}{\pi(y' \mid \theta')}\\
&= \frac{\exp\!\bigl(\theta' h(y)\bigr)\,\pi(\theta')\exp\!\bigl(\theta h(y')\bigr)}
         {\exp\!\bigl(\theta h(y)\bigr)\,\pi(\theta)\exp\!\bigl(\theta' h(y')\bigr)}
\end{aligned}
\end{equation*}
    \EndFor
    \end{algorithmic}
\end{algorithm}

\subsection{Summary Statistics}
The GWESP and GWNSP are defined as:
$$h_{GWESP}(y) = \exp(\tau)\sum_{i=1}^{n-2}\left(1 - \left(1 - \exp(-\tau)\right)^{i}\right)p_{i},$$
$$h_{GWNSP}(y) = \exp(\tau)\sum_{i=1}^{n-2}\left(1 - \left(1 - \exp(-\tau)\right)^{i}\right)np_{i}.$$
$p_{i}$ and $np_{i}$ are the number of connected and non-connected vertex pairs sharing $i$ neighbours, respectively. The decay parameter $\tau$ is set as 0.75 \citep{Lehmann665398}.\\

\subsection{Methods Validity}\label{c_npe_s_method_validity}
In this section, we first discuss the validity of the EM approach. We then align the objective of the NPE with the analytically optimal solution $q^*(\theta_{i})$. Finally, we present the derivation of the M-step updates.\\
A large amount of our proof in this section relies on or is inspired by the foundation built by Jia et al. \citep{jia2023analytically}.

\subsubsection{EM Procedure}
We now give a mathematical proof of the validity of the EM procedure by first considering the ELBO for the joint KL divergence,
$$\text{ELBO}(q)=E_{\Theta_{all}}[log\dfrac{\pi(\Theta_{all},x)}{q(\Theta_{all})}].$$
Under the mean field assumption, the ELBO becomes
\begin{align}
\label{eq:elbo_reformulation}
\text{ELBO}(q)
&= \mathbb{E}_{\theta_g, \Sigma_g}
\left[ \log \pi(\theta_g, \Sigma_g) - \log q(\theta_g, \Sigma_g) \right] \nonumber \\
&\quad + \sum_{i=1}^N \mathbb{E}_{\theta_i}
\Big[ \mathbb{E}_{\theta_g, \Sigma_g}
\big[ \log p(x_i \mid \theta_i) + \log \pi(\theta_i \mid \theta_g, \Sigma_g) \big] \nonumber \\
&\qquad - \log q(\theta_i) \Big].
\end{align}
Note that the above expectations are with respect to $q(\cdot)$.\\ 
We first derive that the optimal $q^{*}(\theta_{i})$ targets the conditional posterior $\pi(\theta_{i}|x_{i},\theta_g, \Sigma_g)$. The derivation logic is to first gather all relevant terms in the expression, and take the functional derivative. By setting it to zero, we can derive the form we want.\\
The relevant terms in the ELBO (Equation~\eqref{eq:elbo_reformulation}) for $q(\theta_i)$ is:
$$\text{ELBO}(q(\theta_i)) = \mathbb{E}_{\theta_i} \left[ \mathbb{E}_{\theta_g, \Sigma_g} \left[ \log p(x_i \mid \theta_i) + \log \pi(\theta_i \mid \theta_g, \Sigma_g) \right] - \log q(\theta_i) \right].$$
We take the functional derivative and set it to zero to achieve the optimality,
$$
\frac{\delta \text{ELBO}(q(\theta_i))}{\delta q(\theta_i)} = \mathbb{E}_{\theta_g, \Sigma_g} \left[ \log p(x_i \mid \theta_i) + \log \pi(\theta_i \mid \theta_g, \Sigma_g) \right] - \log q(\theta_i) - 1 = 0.$$
Rearranging and solving for $q^*(\theta_i)$, we achieve
$$\log q^*(\theta_i) = \mathbb{E}_{\theta_g, \Sigma_g} \left[ \log p(x_i \mid \theta_i) + \log \pi(\theta_i \mid \theta_g, \Sigma_g) \right] + \text{constant},$$
and
$$q^*(\theta_i) = \frac{\exp\left( \mathbb{E}_{\theta_g, \Sigma_g} \left[ \log p(x_i \mid \theta_i) + \log \pi(\theta_i \mid \theta_g, \Sigma_g) \right] \right)}{Z},$$
leaving $Z$ as a normalisation constant. Our derived form is the same as the $q^*$ proportionality expression from Jia et al. derivation as the $\theta_g$ and $\Sigma_g$ are treated as independent of $\theta_i$ \citep{jia2023analytically}.\\
If $q^*(\theta_g, \Sigma_g) = \pi(\theta_g, \Sigma_g | x)$, then $q^*(\theta_i)=\pi(\theta_{i}|\theta_g, \Sigma_g, x_{i})$ also holds in the EM procedure when keeping the global parameters $\theta_g$ and $\Sigma_g$ fixed.\\
Similarly, to demonstrate the optimal $q^*(\theta_g, \Sigma_g)$ targets the posterior $\pi(\theta_g, \Sigma_g|x)$. We pick out terms in the ELBO (Equation~\eqref{eq:elbo_reformulation}) related to global parameters and re-arrange,
$$\text{ELBO}(q(\theta_g, \Sigma_g)) = \mathbb{E}_{\theta_g, \Sigma_g} \left[ \log \pi(\theta_g, \Sigma_g) + \sum_{i=1}^N \mathbb{E}_{\theta_i} [ \log p(x_i, \theta_i \mid \theta_g, \Sigma_g) ] \right].$$
Following the similar optimisation procedure, we can find $q^{*}(\theta_g, \Sigma_g)$ indeed targets the posterior $\pi(\theta_g, \Sigma_g|x)$ with equality holding when $q(\theta_i)$ equals the true conditional posterior.\\
Now we have shown our variational distributions indeed target what they should target, and then we discuss the EM procedure. By sequentially performing the E-step (optimising $q(\theta_i)$) and the M-step (Optimising $q(\theta_g, \Sigma_g)$), the ELBO is guaranteed to monotonically increase (or remain the same at each iteration), which ensures convergence to a (local) maximum.\\
We further discuss some potential limitations in our EM procedure which may hinder our convergence speed. Firstly, we use a sampling procedure to estimate the expectation over our estimated NPE $q(\theta_{i})=q_{\phi}(\theta_{i}|x_{i})$, which may contain bias. However, given the efficiency of sample realisation from the estimated normalising flows, we can generate a large number of samples to avoid the potential bias. Secondly, with the need for a fixed point estimate of $\theta_g$ and $\Sigma_g$ for the E-step, we used MAP from the updated variational distribution from the M-step, which may also cause bias. However, both the variational distribution of $\theta_g$ (Normal) and $\Sigma_g$ (Inverse-Wishart) are regular and well-behaved, meaning that the MAP estimates should serve as adequate point estimates.

\subsubsection{NPE Approximation}
In the previous proof, we have shown the $q^{*}(\theta_{i})$ targets the conditional posterior $\pi(\theta_{i}|x_{i},\theta_g, \Sigma_g)$ with equality holding when keeping the global parameters fixed. The validity of replacing $q^{*}(\theta_{i})$ with the NPE $q_{\phi}(\theta_{i}|x_{i})$ relies on the fact that the objective of both approaches is aligned. With the hierarchical prior $\pi(\theta_{i}|\theta_g, \Sigma_g)$, $q_{\phi}(\theta_{i}|x_{i})$ also targets the conditional posterior $\pi(\theta_{i}|x_{i},\theta_g, \Sigma_g)$.\\
We consider the loss function targeted by the NPE. The derivation logic is that when we minimise for the NPE loss, we are equivalently minimising the KL divergence between our $q_{\phi}(.)$ and $q^*(.)$. Recall the target loss function for NPE is 
$$\mathcal{L}(\phi) = -\mathbb{E}_{\theta_{i}, x_{i}} [ \log q_\phi(\theta_{i} | x_{i}) ].$$
Consider
$$p(x_i, \theta_i) = p(x_i \mid \theta_i)\pi(\theta_i \mid \theta_g, \Sigma_g),$$
noting the global terms are incorporated as the prior terms. Using Bayes' Theorem
$$\pi(\theta_i \mid x_i, \theta_g, \Sigma_g) = \frac{p(x_i \mid \theta_i) \pi(\theta_i \mid \theta_g, \Sigma_g)}{p(x_i \mid \theta_g, \Sigma_g)},$$
moving the right hand side denominator to the left hand side, we get
$$p(x_i, \theta_i) = p(x_i \mid \theta_g, \Sigma_g) \pi(\theta_{i}|x_{i},\theta_g, \Sigma_g).$$
Now consider our NPE loss function again and expand the expectation,
$$\mathcal{L}(\phi) = -\mathbb{E}_{x_i, \theta_i} [ \log q_\phi(\theta_i \mid x_i) ] = -\int p(x_i, \theta_i) \log q_\phi(\theta_i \mid x_i) \, d\theta_i.$$
Inserting $p(x_i, \theta_i)$
$$\mathcal{L}(\phi) = -p(x_i \mid \theta_g, \Sigma_g)\int \pi(\theta_{i}|x_{i},\theta_g, \Sigma_g) \log q_\phi(\theta_i \mid x_i) \, d\theta_i,$$
and noting $ p(x_i \mid \theta_g, \Sigma_g)$ is independent of $\theta_i$ as well as replacing $\pi(\theta_{i}|x_{i},\theta_g, \Sigma_g)$ with $q^{*}(\theta_{i})$ (considering equality in the EM steps),
$$\mathcal{L}(\phi) = -p(x_i \mid \theta_g, \Sigma_g) \cdot \mathbb{E}_{\theta_i} [ \log q_\phi(\theta_i \mid x_i) ].$$
We can see minimisation of NPE loss function is equivalent to minimising $-\mathbb{E}_{\theta_i} [ \log q_\phi(\theta_i \mid x_i) ]$, which is further equivalent to minimising the KL Divergence
$$\text{KL}(q^* \| q_\phi) = \mathbb{E}_{\theta_i} [ \log q^*(\theta_i) - \log q_\phi(\theta_i \mid x_i) ].$$
Note that $\mathbb{E}_{\theta_i} [ \log q^*(\theta_i) ]$ is independent of $\phi$. Therefore, we have rigorously shown the intuitive fact that NPE is a valid approximation of the $q^{*}(\theta_{i})$.\\

\subsubsection{M-step Updates}
In this part, we provide the mathematical derivation of the M-step updates under the Normal-inverse-Wishart prior for $\theta_{g}$ and $\Sigma_{g}$. Recall the optimal solution is of the form
$$q^*(\theta_g, \Sigma_g) \propto \exp\left\{ \mathbb{E}_{\theta_{i}} \left[ \log p(\theta_{i}, \theta_g, \Sigma_g, x) \right] \right\},$$
which can be simplified under the mean-field assumption as:
$$q^*(\theta_g, \Sigma_g) \propto \exp\left\{ \mathbb{E}_{\theta_{i}} \left[ \log \pi(\theta_g, \Sigma_g) + \sum_{i=1}^n \log \pi(\theta_i \mid \theta_g, \Sigma_g) + \sum_{i=1}^n \log p(x_i \mid \theta_i) \right] \right\}.$$
Recognising  \( \log p(x_i \mid \theta_i) \) is independent of $\theta_g$ and $ \Sigma_g$; and independence between $\theta_{i}$, it can be reduced to
\begin{equation}
\label{eq:c_npe_m_step_eq1}
q^*(\theta_g, \Sigma_g) \propto \pi(\theta_g, \Sigma_g) \exp\left\{ \sum_{i=1}^n \mathbb{E}_{\theta_i} \left[ \log \pi(\theta_i \mid \theta_g, \Sigma_g) \right] \right\},
\end{equation}
where we slightly abuse notation by using $\theta_i$ to denote both the joint and individual distributions.\\
From this step onwards, the derivation can be complex, and due to the complexity of the full equations, we will use simplified equations during the derivation process. To prevent it from becoming too abstract, we now explain the derivation logic. This derivation is mainly consolidated from the Jia et al. study \citep{jia2023analytically}, and we build upon their analytical solution to incorporate the Normal-inverse-Wishart prior on $\theta_{g}$ and $\Sigma_{g}$. In their study, they have derived the second component in Equation~\eqref{eq:c_npe_m_step_eq1} will follow a Normal-inverse-Wishart distribution, but they left the left component as 1 (i.e., Uniform). We first re-show their derivations using our own steps. Then we consider this as a problem of multiplication of two Normal-inverse-Wishart distribution to derive our final updates.\\
We first consider $\mathbb{E}_{\theta_i} [ \log \pi(\theta_i \mid \theta_g, \Sigma_g) ]$ in Equation~\eqref{eq:c_npe_m_step_eq1} with the Normal prior structure $\mathcal{N}(\theta_i \mid \theta_g, \Sigma_g)$ and expand the distribution, 
\begin{equation}
\label{eq:c_npe_m_step_eq2}
\mathbb{E}_{\theta_i} \left[ \log \pi(\theta_i \mid \theta_g, \Sigma_g) \right] = \text{Constant} - \frac{1}{2} \log |\Sigma_g| - \frac{1}{2} E_{\theta_i} \left[ (\theta_i - \theta_g)^\top \Sigma_g^{-1} (\theta_i - \theta_g) \right].
\end{equation}

To evaluate the expectation based on the updated variational distribution of $\theta_{i}$, we define
$$ \mu_i = E_{\theta_i}[\theta_i], $$
$$ \Sigma_i = \text{Cov}_{q(\theta_i)}[\theta_i] = E_{\theta_i}\left[ (\theta_i - \mu_i)(\theta_i - \mu_i)^\top \right]. $$
Then, we can expand out the expectation in Equation~\eqref{eq:c_npe_m_step_eq2} using trace trick:
\begin{equation}
\label{eq:c_npe_m_step_eq3}
\mathbb{E}_{\theta_i} \left[ (\theta_i - \theta_g)^\top \Sigma_g^{-1} (\theta_i - \theta_g) \right] = \text{tr}\left( \Sigma_g^{-1} \Sigma_i \right) + (\mu_i - \theta_g)^\top \Sigma_g^{-1} (\mu_i - \theta_g).
\end{equation}
We then insert Equation~\eqref{eq:c_npe_m_step_eq3} back to Equation~\eqref{eq:c_npe_m_step_eq2} and include the summation procedure over all individuals, as in Equation~\eqref{eq:c_npe_m_step_eq1}, 
\begin{equation}
\label{eq:c_npe_m_step_eq4}
\sum_{i=1}^n \mathbb{E}_{\theta_i} \left[ \log \pi(\theta_i \mid \theta_g, \Sigma_g) \right]  = \text{Const} - \frac{n}{2} \log |\Sigma_g| - \frac{1}{2} \sum_{i=1}^{n} \left[ \text{tr}\left( \Sigma_g^{-1} \Sigma_i \right) + (\mu_i - \theta_g)^\top \Sigma_g^{-1} (\mu_i - \theta_g) \right].
\end{equation}
We are now preparing to recognise the expression as our known (Normal-inverse-Wishart) distribution. The logic is that we pick out all terms related to $\theta_{g}$ and completing the square. We expand and rearrange the quadratic terms in Equation~\eqref{eq:c_npe_m_step_eq4},
$$\sum_{i=1}^{n} (\mu_i - \theta_g)^\top \Sigma_g^{-1} (\mu_i - \theta_g) = n \theta_g^\top \Sigma_g^{-1} \theta_g - 2 \theta_g^\top \Sigma_g^{-1} \sum_{i=1}^{n} \mu_i + \sum_{i=1}^{n} \mu_i^\top \Sigma_g^{-1} \mu_i.$$
By defining:
$$\bar{\mu} = \frac{1}{n} \sum_{i=1}^{n} \mu_i,$$
and complete the squares, we can get
$$\begin{aligned}
\sum_{i=1}^{n} (\mu_i - \theta_g)^{\top} \Sigma_g^{-1} (\mu_i - \theta_g) &= n (\theta_g - \bar{\mu})^{\top} \Sigma_g^{-1} (\theta_g - \bar{\mu}) + \sum_{i=1}^{n} (\mu_i - \bar{\mu})^{\top} \Sigma_g^{-1} (\mu_i - \bar{\mu})
\end{aligned}.$$
We insert this back to Equation~\eqref{eq:c_npe_m_step_eq4} to consider all relevant terms, the Equation~\eqref{eq:c_npe_m_step_eq4} then becomes
\begin{equation}
\label{eq:c_npe_m_step_eq5}
\begin{aligned}
\sum_{i=1}^n \mathbb{E}_{q(\theta_i)} \left[ \log \pi(\theta_i \mid \theta_g, \Sigma_g) \right]  &= \text{Const} - \frac{n}{2} \log |\Sigma_g| - \frac{n}{2} (\theta_g - \bar{\mu})^{\top} \Sigma_g^{-1} (\theta_g - \bar{\mu}) \\
&\quad - \frac{1}{2} \sum_{i=1}^{n} \text{tr}(\Sigma_g^{-1} \Sigma_i) - \frac{1}{2} \sum_{i=1}^{n} (\mu_i - \bar{\mu})^{\top} \Sigma_g^{-1} (\mu_i - \bar{\mu}).
\end{aligned}
\end{equation}
We can already see the Normal part of the Normal-inverse-Wishart. Since the remaining procedures are quite standard, we jump a bit and omit some naive explanations. We combine the last two terms in Equation~\eqref{eq:c_npe_m_step_eq5}:
$$\sum_{i=1}^n \left[ \operatorname{tr}\left( \Sigma_g^{-1} \Sigma_i \right) + (\mu_i - \bar{\mu})^\top \Sigma_g^{-1} (\mu_i - \bar{\mu}) \right] = \operatorname{tr} \left( \Sigma_g^{-1} \Psi_1 \right),$$
with $\Psi_1 = \sum_{i=1}^n \left[ \Sigma_i + (\mu_i - \bar{\mu})(\mu_i - \bar{\mu})^\top \right]$.
We can identify the Normal-Inverse-Wishart for the exponential terms in Equation~\eqref{eq:c_npe_m_step_eq5}
$$\exp\left\{ \sum_{i=1}^n \mathbb{E}_{q(\theta_i)} \left[ \log \pi(\theta_i \mid \theta_g, \Sigma_g) \right] \right\} = \text{NIW}\left( \theta_g, \Sigma_g \mid \mu_{1}, k_{1}, \Psi_{1}, \nu_{1} \right),$$
with
$$\mu_{1}=\bar{\mu} = \frac{1}{n} \sum_{i=1}^{n} \mu_i,$$
$$k_{1}=n,$$
$$\Psi_{1} = \sum_{i=1}^{n} \left[ \Sigma_i + (\mu_i - \bar{\mu})(\mu_i - \bar{\mu})^{\top} \right],$$
$$\nu_{1} = n - d - 2.$$
Our derived $\Psi_{1}$ is an equivalent expression to $\Psi_{1} = \sum_{i=1}^{n} \mathbb{E}[\theta_i \theta_i^\top] - n \bar{\mu} \bar{\mu}^\top$ as in Jia et al. paper \citep{jia2023analytically}. \\
We then proceed to consider the Equation~\eqref{eq:c_npe_m_step_eq1} $q^*(\theta_g, \Sigma_g) \propto \pi(\theta_g, \Sigma_g) \exp\left\{ \sum_{i=1}^n \mathbb{E}_{q(\theta_i)} \left[ \log \pi(\theta_i \mid \theta_g, \Sigma_g) \right] \right\}$ by inputting the Normal-inverse-Wishart prior $\pi(\theta_g, \Sigma_g) \sim \text{NIW}\left( \theta_g, \Sigma_g \mid \mu_{0}, k_{0}, \Psi_{0}, \nu_{0} \right)$.\\
We treat this problem as a multiplication of two Normal-inverse-Wishart distributions:
$$p_1(\theta_g, \Sigma_g) = \mathcal{N}\left( \theta_g \mid \mu_0, \frac{\Sigma_g}{\kappa_0} \right) \cdot \mathcal{IW}\left( \Sigma_g \mid \nu_0, \Psi_0 \right),$$
$$p_2(\theta_g, \Sigma_g) = \mathcal{N}\left( \theta_g \mid \mu_1, \frac{\Sigma_g}{\kappa_1} \right) \cdot \mathcal{IW}\left( \Sigma_g \mid \nu_1, \Psi_1 \right).$$
Equation~\eqref{eq:c_npe_m_step_eq1} then becomes the problem we want to target and reduce,
$$q^*(\theta_g, \Sigma_g) = p_1(\theta_g, \Sigma_g) \cdot p_2(\theta_g, \Sigma_g),$$
where $p_1(\theta_g, \Sigma_g)$ represents the Normal-inverse-Wishart prior, and $p_2(\theta_g, \Sigma_g)$ represents the derived Normal-inverse-Wishart distribution in Equation~\eqref{eq:c_npe_m_step_eq5}.\\
We can then recognise the distribution for $q^{*}(\Sigma_g)$,
$$q^{*}(\Sigma_g) \sim IW(\Psi_{n},\nu_{n}),$$
where
$$\nu_{n} = \nu_0 + \nu_1 + d + 2,$$
$$\Psi_{n} = \Psi_0 + \Psi_1 + \frac{\kappa_0 \kappa_1}{\kappa_0 + \kappa_1} (\mu_0 - \mu_1)(\mu_0 - \mu_1)^\top.$$
Then, $q^{*}(\theta_g|\Sigma_g)$ becomes,
$$q^{*}(\theta_g|\Sigma_g) \sim \mathcal{N}(\mu_{n},k_{n}),$$
where
$$\mu_{n} = \frac{\kappa_0 \mu_0 + \kappa_1 \mu_1}{\kappa_0 + \kappa_1},$$
$$k_{n} = \kappa_0 + \kappa_1.$$
We insert in the two Normal-inverse-Wishart distributions to obtain the variational updates.
$$\nu_{n} = \nu_0 + n,$$
$$\Psi_{n} = \Psi_0 + \sum_{i=1}^{n} \left[ \Sigma_i + (\mu_i - \bar{\mu})(\mu_i - \bar{\mu})^{\top} \right] + \frac{\kappa_0 n}{\kappa_0 + n} (\mu_0 - \bar{\mu})(\mu_0 - \bar{\mu})^\top,$$
$$\mu_{n} = \frac{\kappa_0 \mu_0 + \sum_{i=1}^{n}\mu_{i}}{\kappa_0 + n},$$
$$k_{n} = \kappa_0 + n.$$

\printbibliography

\end{document}